# A Power Efficient Artificial Neuron Using Superconducting Nanowires

Emily Toomey[1], Ken Segall[2], Karl K. Berggren*[1]

[1]*Massachusetts Institute of Technology, Department of Electrical Engineering and Computer Science, Cambridge, MA 02139, USA*

[2]*Colgate University, Department of Physics, Hamilton, NY 13346, USA*

*contact: berggren@mit.edu

**ABSTRACT**

With the rising societal demand for more information-processing capacity with lower power consumption, alternative architectures inspired by the parallelism and robustness of the human brain have recently emerged as possible solutions. In particular, spiking neural networks (SNNs) offer a bio-realistic approach, relying on pulses analogous to action potentials as units of information. While software encoded networks provide flexibility and precision, they are often computationally expensive. As a result, hardware SNNs based on the spiking dynamics of a device or circuit represent an increasingly appealing direction. Here, we propose to use superconducting nanowires as a platform for the development of an artificial neuron. Building on an architecture first proposed for Josephson junctions, we rely on the intrinsic nonlinearity of two coupled nanowires to generate spiking behavior, and use electrothermal circuit simulations to demonstrate that the nanowire neuron reproduces multiple characteristics of biological neurons. Furthermore, by harnessing the nonlinearity of the superconducting nanowire's inductance, we develop a design for a variable inductive synapse capable of both excitatory and inhibitory control. We demonstrate that this synapse design supports direct fanout, a feature that has been difficult to achieve in other superconducting architectures, and that the nanowire neuron's nominal energy performance is competitive with that of current technologies.

## 1. INTRODUCTION

The human brain has long been a subject of fascination due to the wide variety of complex operations made possible by groups of a single component—the neuron. Now, as computation needs are rapidly approaching the limits of traditional von Neumann architectures, the neuron's unique features have led it to become a source of inspiration for new directions for the advancement of computing. Unlike conventional computing schemes, the human brain benefits from characteristics such as extensive parallelism and robustness to errors, allowing it to operate efficiently despite slow speeds on the order of a few Hz. These appealing qualities have spurred the development of technologies that use the brain as a platform for information processing, ranging from small scale modeling of single neuron dynamics to large-scale parallel computing.

At the heart of this concept are Spiking Neural Networks (SNNs), which seek to mimic the spiking dynamics of the brain in order to encode information, with the additional benefit of possibly offering new insight into the brain's functionality. In this scheme, spikes serve as the tokens of information, while neurons act analogously to logic gates, producing a single output in response to a combination of multiple inputs [1]. Past approaches to SNNs vary both in degree of bio-realism and in how the spikes are implemented. While software approaches that hard-code spiking dynamics offer flexibility and precision, they are computationally expensive. As a result, other SNNs have used a hardware implementation, relying on devices with intrinsic dynamics that replicate neuron behavior as a means of reducing computation costs. Hardware approaches have been explored in a wide variety of platforms, including CMOS [2], magnetic materials [3], memristors [4] [5], and superconducting Josephson junctions (JJs) [6] [7]. CMOS circuits provide large-scale integration, but are heavy on power dissipation and need many components to achieve biological realism. Memristors and magnetic materials have device characteristics which emulate neural dynamics, but can be slow and are also not energy efficient. Josephson junctions are a fast and energy-efficient technology, but need more components in large networks due to poor fan-out/fan-in properties; furthermore, their weak action potentials are unable to be viewed directly, making diagnostics difficult.

These developing technologies highlight several characteristics that are critical for an artificial neuron: (1) inherent device dynamics that are capable of producing spiking behavior; (2) tunable synapses between neurons to allow for the expansion into larger networks with adjustable connectivity; and (3) low power dissipation, both in the dynamic firing state of the neuron and in the static state. As has been emphasized in prior literature[2], optimizing static power dissipation is particularly vital to the goal of creating an energy efficient network.

Here we propose an artificial neuron based on superconducting nanowires operating at cryogenic temperatures. Superconductors are prime candidates for generating low-power spiking behavior due to their inherent nonlinearity and negligible static power dissipation. Building on an architecture first implemented in Josephson junctions [6], we rely on the coupling between two shunted nanowires to act analogously to a two-channel neuron for an action potential. We start by describing the nonlinear dynamics of superconducting nanowires, and then present the architecture of the nanowire-based neuron. Using electrothermal circuit simulations, we demonstrate that the device is able to replicate several behaviors of a single neuron. Finally, we present a synapse design and discuss the advantages of the nanowire neuron, including fanout and an energy figure of merit two orders of magnitude better than that of competing technologies, which are critical in moving towards the parallelism of the human brain.

## 2. THE NANOWIRE NEURON MODEL

Superconducting nanowires possess an inherent nonlinearity that serves as the building block of our artificial neuron model. We begin by briefly describing this nonlinearity in a single nanowire, and then present the nanowire neuron circuit and its basic operation principles.

### 2.1 Relaxation Oscillations

The intrinsic nonlinearity of superconducting nanowires makes them ideal candidates for the hardware generation of spiking behavior. When a bias current flowing through a superconducting nanowire exceeds a threshold known as the critical current ($I_c$), superconductivity breaks down and the nanowire becomes resistive, producing a voltage. The nanowire only switches back to the superconducting state once the bias current is reduced below a level called the retrapping current ($I_r$), and the resistive portion (the "hotspot") cools down. When the nanowire is placed in parallel with a shunt resistor, this switching process participates in electrothermal feedback with the shunt, producing relaxation oscillations[8].

Relaxation oscillations can be viewed in the context of a simplified action potential. Like the Na+ influx and K+ outflux currents of a neuron, the influx and outflux currents from the nanowire to the shunt resistor are governed by different timescales, $\tau_1$ and $\tau_2$. As shown in Figure 1a, the rising edge of the output voltage is defined by $\tau_1=L/(R_s+R_{hs})$, where $L$ is the inductance of the nanowire, $R_s$ is the shunt resistance, and $R_{hs}$ is the resistance of the nanowire hotspot, usually on the order of ~ 1-10 k$\Omega$. Conversely, the outflux current occurs when the nanowire is no longer resistive and the bias current is redirected from the shunt; this reduced resistance results in a slower time constant $\tau_2= L/R_s$ which defines the falling edge of the output voltage. For typical nanowires devices, $\tau_1$ ~ 100 ps and $\tau_2$ ~ 1 ns. The two currents are "gated," as shown by the insets in Fig. 1a, by the state of the nanowire—when the state is resistive (producing a voltage), the influx current flows into the shunt, and when it is superconducting, the outflux current flows back to the nanowire.

The inductance of superconducting nanowires is dominated by an intrinsic material property known as kinetic inductance [9], and is defined per unit length of the structure. As a result, it is possible to tune these time constants by changing the length of the nanowire, with a longer wire leading to a higher inductance and thus a longer timescale. Figure 1b shows a scanning electron micrograph of a typical superconducting nanowire with a meandering geometry designed for maximizing the total device inductance. An example of experimentally observed relaxation oscillations for such a device is displayed in Figure 1c.

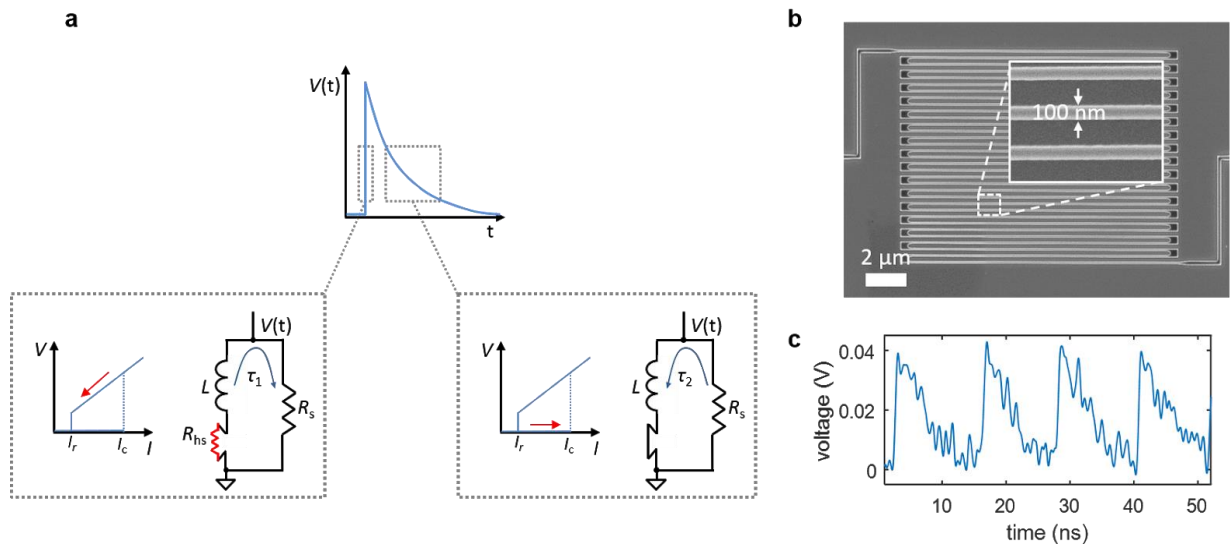

**Figure 1: Relaxation oscillations in superconducting nanowires, which serve as the foundation of the nanowire neuron's spiking behavior.** (a) Simplified model of a relaxation oscillation. The output voltage is defined by two time constants. The rising edge occurs when the superconducting nanowire has switched into the resistive state ($R_{hs} > 0$), and the bias current is redirected to the shunt resistor. The falling edge takes place after the nanowire regains superconductivity ($R_{hs} = 0$), and the bias current is redirected from the shunt resistor. (b) Example of a superconducting nanowire in a long, meandered design for obtaining a high kinetic inductance. (c) Experimentally measured relaxation oscillations in a long superconducting nanowire shunted by 50 $\Omega$.

## *2.2 The Neuron Model*

Although a shunted nanowire on its own produces oscillations analogous to action potentials, as the bias current increases, the output signal eventually accumulates a voltage offset. This effect deviates from true neuron behavior, as the cell must maintain a constant resting potential (approximately -70 mV). To overcome this difference, we have implemented a neuron architecture based on one that was first proposed for Josephson junctions [6], as shown in Figure 2. The circuit consists of two shunted nanowires—the main oscillator and the control oscillator—linked together in a superconducting loop. A bias current $I_{bias}$ is applied to both oscillators such that they are each biased right below their critical currents, but in opposite directions. To trigger an action potential, a small input current pulse $I_{in}$ (Fig. 2a) is applied and sums with the bias current to exceed $I_c$ of the main oscillator, causing it to switch (Fig. 2d). The control does not fire since the input opposes the direction of its bias.

Once the main oscillator switches, current is added to the superconducting loop in the counterclockwise direction (Fig. 2b), which sums with the bias current to fire the control oscillator (Fig. 2c). The control oscillator removes counterclockwise current from the loop, allowing the main oscillator to fire again. Without the presence of the control oscillator, the main oscillator would only be able to fire once, since the counterclockwise current added to the loop would reduce the total current through the nanowire of the main oscillator below its $I_c$. The voltage from the main oscillator node (Fig. 2e) serves as the spiking output that is carried down to the next neuron via the synapse. Unlike the output from the single shunted nanowire, the output of the two-nanowire circuit does not accumulate a bias offset, making it a suitable spiking signal.

In the context of the two-channel neuron model, the main oscillator acts analogously to the Na+ influx current by adding flux to the superconducting loop in the form of a circulating current. The control oscillator acts analogously to the K+ outflux current by reducing the circulating current, resetting the neuron and allowing the main oscillator to fire again. As described in Ref. [8], the rate at which each oscillator fires depends on the magnitude of the bias current, paralleling the voltage-dependent rate constants of ion gates in the Hodgin-Huxley model [10].

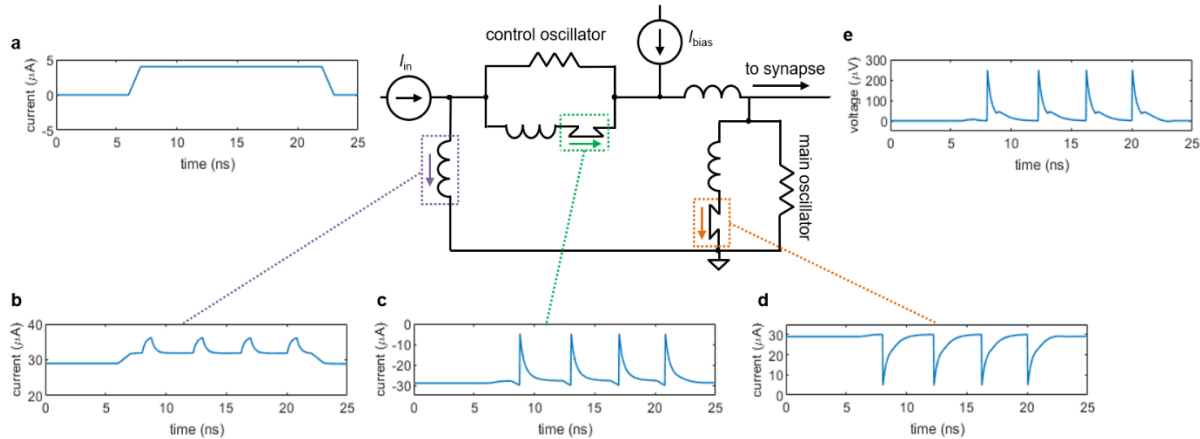

**Figure 2: Circuit simulations of the two-nanowire soma, where the two oscillators act analogously to the two ion channels in the simplified neuron model.** (a) Input pulse, $I_{in} = 4$ µA. (b) Current through the loop inductor. (c) Current through the control nanowire. The control nanowire reduces the amount of counter-clockwise current circulating in the loop, allowing the main nanowire to fire again. (d) Current through the main nanowire. (e) Output voltage pulse that is sent to the synapse. For these simulations, the critical current of the control nanowire is $I_{c,control} = 30$ µA, the critical current of the main nanowire is $I_{c,main} = 30$ µA, and $I_{bias} = 58.6$ µA.

## 3. SINGLE NEURON CHARACTERISTICS

Neurons display a wide variety of traits unique to certain populations, allowing them to collectively interact to achieve varied and complex tasks. While no single neuron possesses all possible traits, the basic functionality of an artificial neuron can be evaluated by demonstrating some common bio-realistic characteristics. Here we present multiple neuron behaviors that can be achieved with the nanowire neuron, using electrothermal circuit simulations conducted in LTSpice. The simulations implement material-specific characteristics and nanowire hotspot dynamics as described in previous literature [11] [12], and have been shown to reliably reproduce experimental data pertaining to nanowire relaxation oscillations [8].

*3.1 Threshold Response*

A general characteristic of biological neurons is their inability to fire unless the input signal exceeds a certain threshold. Figure 3a shows the threshold voltage response of the nanowire neuron when the bias current is held constant and the input current is varied. As evident in the plot, the neuron does not begin firing until the input current passes a threshold, defined by when the sum of the bias and input current through the main nanowire exceed its $I_c$. Above the threshold, the peak voltage of the spike output is essentially constant. However, as emphasized by Izhikevich [13], biological neurons have a threshold that may be varied by previous activity, such as an inhibiting input that reduces it. Figure 3b and c illustrate this process ("threshold variability") in the nanowire neuron; an initial subthreshold input pulse (Fig. 3b) fails to elicit a spike, while a later input pulse of the same magnitude triggers a spike (Fig. 3c) after a smaller negative (inhibitory) pulse reduces the firing threshold. It should be noted that when the preceding pulse was of the opposite polarity (excitatory), no spike was triggered. This behavior is consistent with the expectations of Ref. [13].

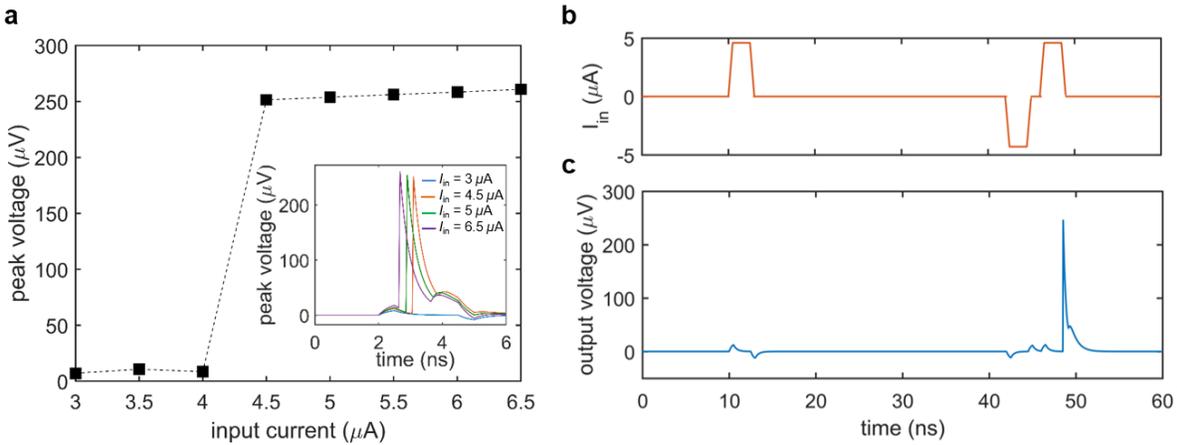

**Figure 3: Firing threshold of the two-nanowire neuron.** (a) Peak output voltage as a function of input current under a constant bias ($I_{bias}$ = 58.6 µA). The plot illustrates that the neuron does not spike until the input current exceeds 4 µA, at which point it fires with output voltages of the same amplitude. Inset shows the time domain voltage output of the neuron for different input currents. (b) Input to the neuron, leading to a reduction in firing threshold by a preceding inhibitory pulse. (c) Spiking output of the neuron in response to the inputs of (b), demonstrating that the nanowire neuron's firing threshold is variable. For this simulation, $I_{bias}$ = 57.62 µA, the excitatory inputs $I_{in}$ = 4.6 µA, and the inhibitory input $I_{in}$ - = -4.3 µA.

*3.2 Refractory Period*

In addition to exhibiting a firing threshold, the nanowire neuron displays a refractory period, which we define to be the minimum time between two input pulses such that both pulses elicit a spike. Figure 4 illustrates this response. When two pulses are separated enough in time so that the main oscillator is biased close to its critical current when the second input pulse arrives, then the second pulse will cause a spike (Fig. 4a). However, if the second pulse arrives before the bias current has fully returned to the main oscillator, then the sum of the second input pulse and the bias will not be sufficient to switch the nanowire and trigger the neuron (Fig. 4b). As a result, the refractory period is limited by the time it takes to fully bias the main oscillator again.

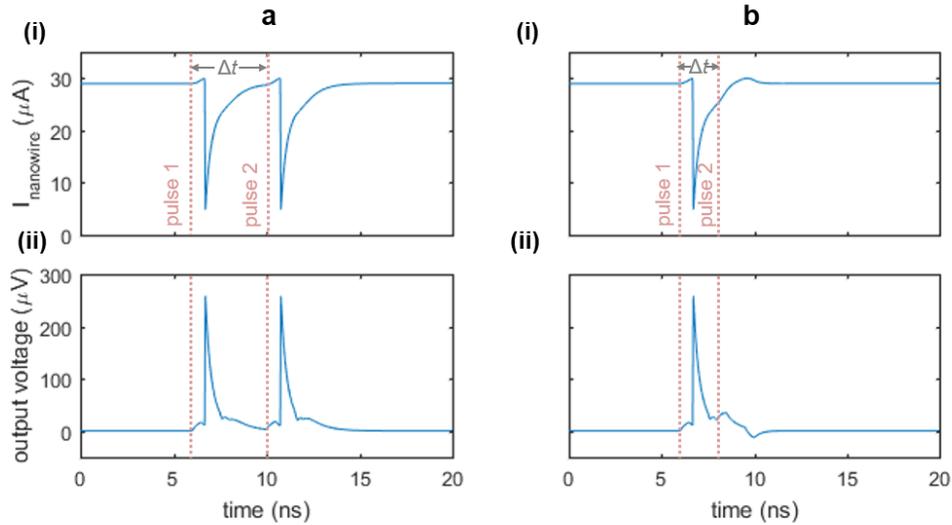

**Figure 4: Refractory period of the two-nanowire neuron.** (a) Response when there is sufficient time between two inputs to each elicit a separate spike. Parameters: $I_{bias} = 58$ µA, $I_{in} = 6$ µA, $\Delta t = 4$ ns. The pink dashed lines indicate the beginning of the rising edge of each pulse. (b) Response when there is insufficient time between two input pulses, causing the neuron to fire only once. Parameters are the same as in (a), except $\Delta t = 2$ ns. For both cases, panel (i) displays the current through the nanowire of the main oscillator, while panel (ii) displays the output voltage of the neuron.

### 3.3 Near-Coincidence Detection

The neuron may also be under-biased such that two pulses in rapid succession are required to elicit a spike. Figure 5 displays the response of the nanowire neuron to two input pulses with different time delays $\Delta t$ between them. In this case, in order for the second pulse to fire the neuron, the delay must be less than 3 ns, as shown in Fig. 5b. This behavior is similar to the response of integrating neurons to high frequency inputs [13], and demonstrates that the nanowire neuron may be used for near-coincidence detection of pulses, depending on its biasing conditions.

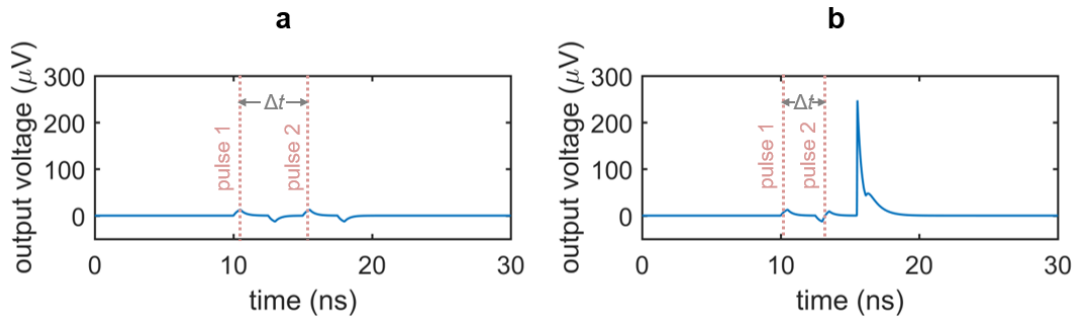

**Figure 5 : Near-coincidence detection of pulses.** Output voltage of the neuron when the time between successive input pulses $\Delta t$ is a) 5 ns b) 3 ns. The pink dashed lines indicate the rising edge of each pulse. Two pulses must be in rapid succession in order to fire the under-biased neuron, demonstrating that it may be used for near-coincidence detection of input pulses. Parameters: $I_{in} = 4.6$ µA, $I_{bias} = 57.62$ µA.

## 3.4 Class I Behavior

Biological neurons differ in their response to varying signal strengths. Whereas Class I neurons have a spiking frequency that increases with increasing input strength, Class II neurons maintain a constant firing rate[13] [6]. Figure 6 illustrates the spiking behavior of the nanowire neuron at different levels of bias current. Fig. 6a shows the time-domain voltage output of the neuron as the bias current is increased, and suggests an increase in spiking frequency. This response is confirmed by observing the voltage output's frequency spectrum displayed in Fig. 6b, which shows a shift in the spiking frequency to higher levels with increasing bias. Consequently, the nanowire neuron has Class 1 behavior. The modulation of spiking frequency by bias current demonstrates that the frequency of the nanowire neuron output may be used to glean information about its input conditions.

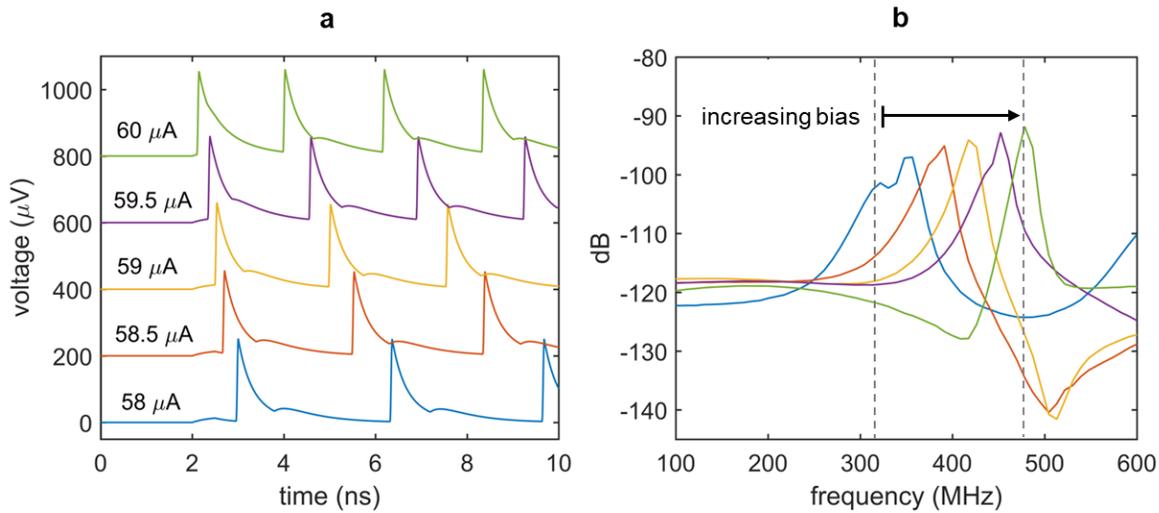

**Figure 6: Effect of bias current on spiking frequency.** (a) Time domain simulations of the two-nanowire neuron with different bias currents. $I_{in}$ = 6 µA for all simulations. Traces have been shifted from one another in the y-axis for clarity. (b) Fourier transform of the voltage output for each biasing condition. The shift in peak frequency with bias current indicates that the circuit acts like a Class I neuron.

## 3.5 Parabolic Bursting

Some neurons, such as thalamic and dopaminergic neurons [14], display a unique mode of behavior called bursting, in which the cell alternates between the resting and firing states. The transition between states may be dictated by slow changes in low levels of intracellular calcium ions, which influence the conductance of K+ [15]. As a result, a small, slowly varying signal controls the rapid dynamics of the action potential, leading to alternating periods of resting and firing. This process is considered to be an important aspect of electrical activity in the brain.

Due to the significance of bursting in biological neurons, past work has sought to replicate similar behavior in platforms such as digital silicon models [16] and Josephson junction models [17] by injecting a low-frequency ac signal into the system. Figure 7 shows the result when a similar technique is applied to the nanowire neuron. In this case, a weak ac signal ($f$ = 50 Hz, $I_{ac}$ = 4 µA) shown by the dashed red line in Fig. 7a is coupled into the bias port of the neuron, causing the neuron to alternate between the resting and firing states, as reflected in the resulting output voltage signal. A close examination of the timing between adjacent spikes (see Fig. 7b) shows that the spiking frequency increases and then decreases over the firing period, a phenomenon known as parabolic bursting[18]. This behavior was first observed in neuron R15 of the abdominal ganglion of Aplysia [19] [20], and has since been demonstrated in many other cells. The ability of the nanowire neuron to replicate similar dynamics may be therefore be useful for performing a wider range of functions.

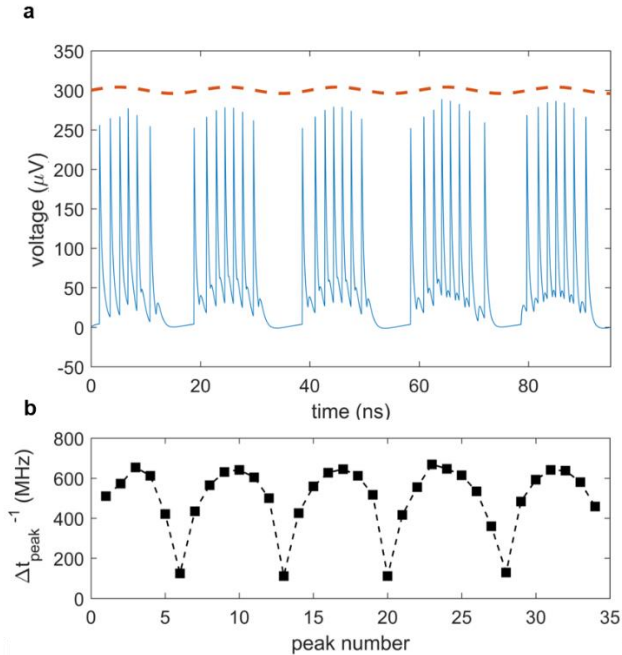

**Figure 7: Parabolic bursting in the two-nanowire neuron.** (a) Output voltage of the two-nanowire neuron when the bias is coupled to a weak sinusoidal drive ($f$ = 50 MHz, $I_{ac}$ = 4 µA). The red dashed curve indicates the sinusoidal drive, shifted in the y-axis for clarity. (b) The inverse of the time between adjacent peaks shows that the time difference follows a parabolic form. Parameters: $I_{bias}$ = 59 µA, $I_{in}$ = 6 µA.

### 3.6 Axon: Transmission Line Characteristics

After an action potential occurs in a biological neuron, the output signal propagates down the axon as if sent through a delay line [1]. This delay is valuable in that it preserves time domain information, potentially facilitating behaviors that rely on the recognition of specific spatio-temporal patterns [13]. Such pattern recognition is often not possible in SNNs with traditional wiring, since signals travel too rapidly for timing information to be maintained [1]. This is not the case in superconducting transmission lines. Recent work has shown that the high kinetic inductance of superconducting transmission lines, like those made out of niobium nitride, results in propagation speeds of ~2% $c$, where $c$ is the speed of light in vacuum [21]. As illustrated in Figure 8, a simulated nanowire neuron output sent through a superconducting transmission line model [22] is delayed by ~100-500 ps, close to the full width of an action potential. In mammalian brains, the full width of the action potential is typically a few milliseconds [23], and so by analogy the axonal delay shown in Figure 8 would also be on the millisecond timescale. This delay would be on the order of typical delays in cortical systems, for example the cortico-cortical delay in mammals [24]; if longer delays are needed, the transmission line can simply be made longer.

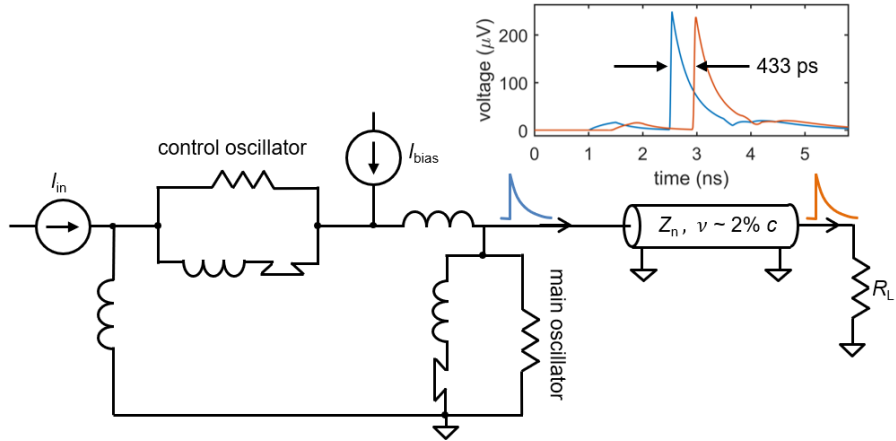

**Figure 8: A superconducting transmission line as an axon.** Simulations of a superconducting transmission line show that the spikes can be delayed on the order of ~0.5 ns, depending on the length of the structure. This could enable the storage of timing information in addition to frequency information. Transmission line parameters: nanowire inductance $L_n = 0.3$ nH/µm, nanowire capacitance $C_n = 0.1$ fF/µm, propagation speed $v = 1.9\%c$, transmission line length $l = 2.5$ mm. Shorter transmission lines on the order of 800 um still had delays of ~140 ps.

## 4. THE SYNAPSE

The collective dynamics of a neural network depend on the ability of a neuron to influence the behavior of another downstream neuron via a synapse. Here we introduce an inductive synapse that can be integrated with the nanowire neuron to facilitate downstream control. We start by demonstrating excitatory and inhibitory control, and then present a scheme for tuning the synaptic strength.

### 4.1 The Inductive Synapse

Figure 9 illustrates the circuit schematic of an inductive synapse that may be implemented in the nanowire neuron. Similar to the slow release of neurotransmitters in response to an action potential, the inductive synapse relies on the slow charging of a large inductor in response to the nanowire neuron's more rapid voltage spikes. The energy stored in the large synapse inductor is then discharged as current into the input port of the target neuron, modulating its behavior.

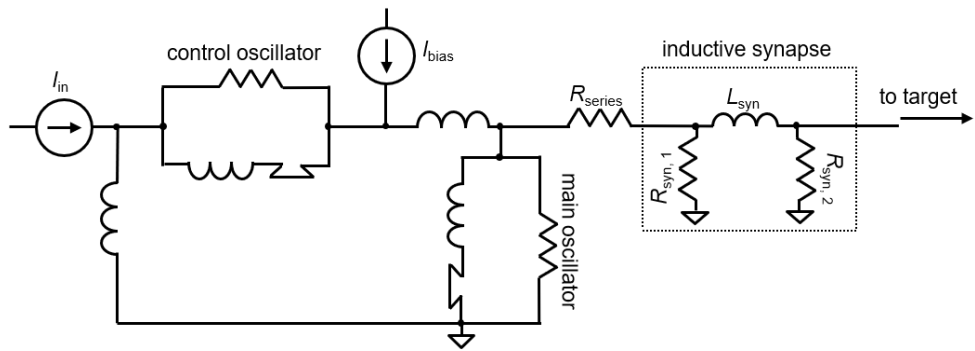

**Figure 9: Circuit schematic of the nanowire neuron with an inductive synapse.** The output voltage of the neuron charges the large synapse inductor $L_{syn}$. The inductor discharges current into the input port of the downstream target neuron, modulating its behavior. A series resistor $R_{series}$ is in place to reduce backaction into the main neuron.

The effect of the inductive synapse can be either excitatory or inhibitory based on the sign of the bias current applied to the upstream (main) neuron. These two cases are shown in Figure 10, which displays the voltage outputs of the main and target neurons, as well as the current through the synapse inductor. In the excitatory case (Fig. 10a), the upstream neuron is positively biased such that the inductive synapse discharges a positive current, causing an under-biased target neuron to fire. In the inhibitory case (Fig. 10b), the upstream neuron is negatively biased, causing a negative current to be sent from the synapse to the target, turning off a target that was biased in the firing state. Consequently, the inductive synapse allows for both types of control between neurons.

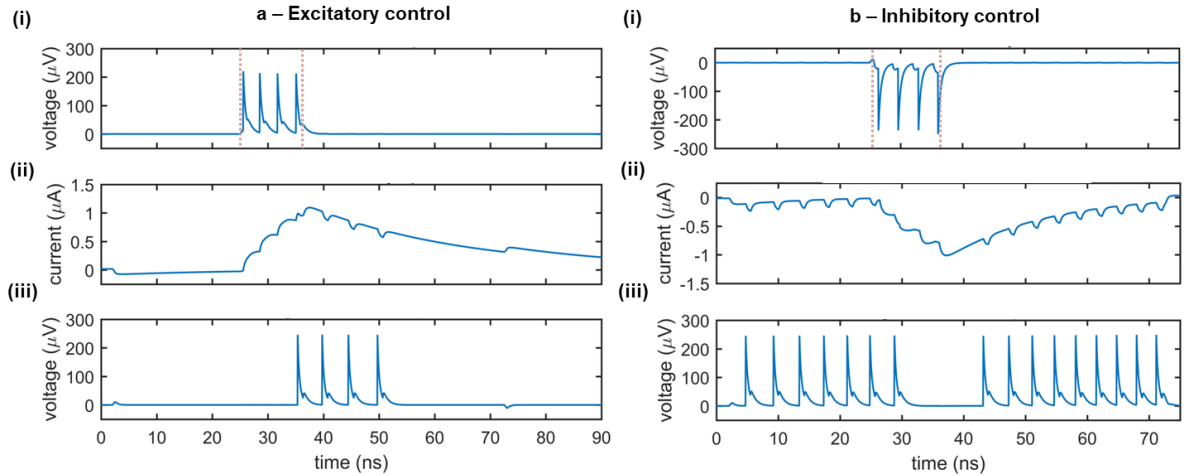

**Figure 10: Downstream control of the inductive synapse.** (a) Excitatory control. Parameters: $I_{bias,main} = 59$ µA, $R_{series} = 14$ Ω, $R_{syn,1} = 40$ Ω, $L_{syn} = 0.265$ µH, $R_{syn,2} = 40$ Ω, $I_{bias,target} = 57.17$ µA, $I_{in} = 4.6$ µA. (b) Inhibitory control. Parameters $I_{bias,main} = -58.6$ µA, $R_{series} = 24$ Ω, $R_{syn,1} = 45$ Ω, $L_{syn} = 0.23$ µH, $R_{syn,2} = 40$ Ω, $I_{bias,target} = 57.68$ µA, $I_{in} = 4.6$ µA. For both cases: Panel (i) displays the output voltage of the main neuron, with the red dashed lines indicating the rising and falling edge of the input signal; Panel (ii) displays the current through the synapse inductor; Panel (iii) displays the output voltage of the downstream target neuron.

## 4.2 Variable Strength Synapse

Although the inductive synapse of Figure 10 can be engineered for both excitatory and inhibitory control of a downstream neuron, a fundamental property of artificial neural networks is the ability to modulate that control by adjusting synaptic strength. One possible scheme for implementing such variability in the inductive synapse is to incorporate superconducting nanowires as a different circuit element: a tunable inductor. A nanowire's kinetic inductance increases with increasing bias current, reaching an enhancement of 10-20% near $I_c$ [25]. This modulation has been incorporated into the circuit model of the superconducting nanowire used in these simulations [12]. By placing a high inductance nanowire with an ideal current source in parallel with the synapse inductor, the overall parallel inductance of the synapse can be modulated, which in turn changes the amount of current sent to the target neuron. Figure 11 shows the simulated results for the case of an inhibitory synapse. When a higher modulating current $I_{mod}$ is applied to the nanowire inductor, the overall parallel inductance increases, reducing the amount of current sent to the target. It is important to note that the polarity of the modulating current is not important, since the change in kinetic inductance depends only on the magnitude of the modulating current in relation to its $I_c$. For instance, Figure 11b illustrates that the modulation in synaptic current for $I_{mod} = 5$ µA and $I_{mod} =$

-5 µA is roughly the same. As a result, it is clear that the modulation is due to the change in kinetic inductance, and not simply an injection of current by $I_{mod}$ in the opposing direction.

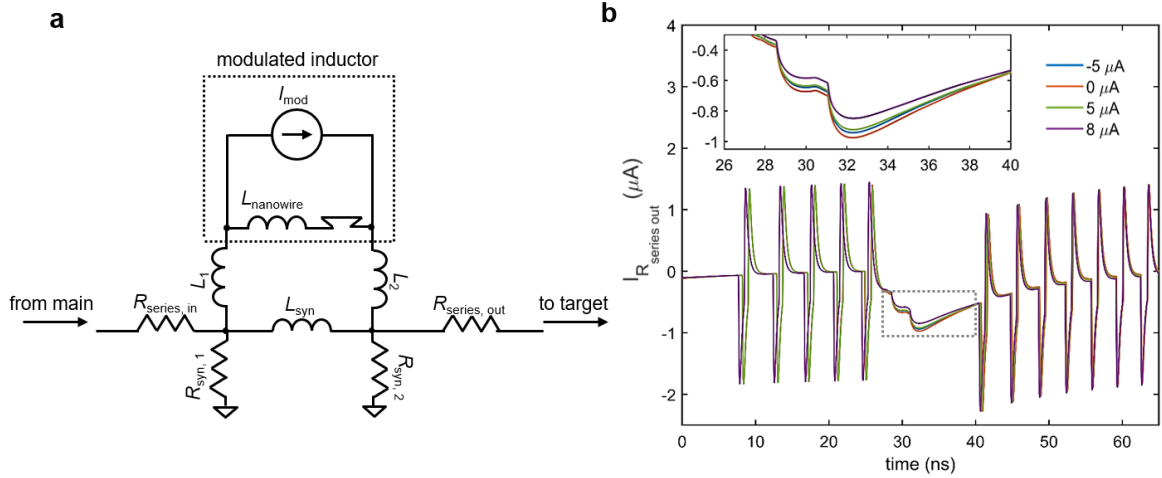

**Figure 11: Modulating the inductive synapse.** (a) Circuit schematic of the inductive synapse with a high-inductance nanowire and ideal current source placed in parallel. $L_1$, $L_2 \ll L_{nanowire}$, $L_{syn}$. $R_{series, out}$ has been added to further prevent backaction from the target neuron. (b) Simulation of the current through $R_{series,out}$ in an inhibitory synapse as a function of different modulation currents. Spikes represent backaction from the firing of the target neuron. (Inset) Enlarged view of the boxed area. Higher modulation currents increase the overall synapse inductance, reducing the amount of current sent to the target. The +/- 5 µA modulation currents have nearly the same effect, showing that modulation is not polarity-dependent. Parameters: $R_{series,in} = 25$ Ω, $R_{syn,1} = 39$ Ω, $R_{syn,2} = 40$ Ω, $R_{series,out} = 0.1$ Ω, $L_{syn} = 0.45$ µH, $L_{nanowire} = 0.275$ µH, $I_{c, nanowire} = 6$ µA, $L_1 = L_2 = 50$ pH, , $I_{bias,target} = 57.65$ µA, $I_{bias,main} = -59.5$ µA

    In conjunction with the plasticity of synapses, the high degree of parallelism in the brain creates a densely connected, adaptable network able to optimize and adjust for different conditions. An example of fanout in the nanowire neuron is shown in Figure 12. As shown in Fig. 12a, a single neuron is connected to four target neurons through four separate tunable synapses. When each of the four modulating currents is set to $I_{mod} = 0$, the firing of all four targets is inhibited by the main neuron, as illustrated in Fig. 12b. To weaken the connection with one of the targets (target #4), $I_{mod,4}$ is set to 8 µA, turning off the inhibiting action on target #4 but allowing it to remain on the three other target neurons, as shown in Fig. 12c. Comparison of the synaptic currents for each of the four targets shows that the synaptic current for target #4 is reduced as a result of the modulated inductance. This modulation illustrates that the nanowire neuron is able to be used in a parallel network where the strength of individual synaptic connections can be adapted.

    Fan-out may be one area where nanowires will be an improvement over Josephson junctions [6], another superconducting technology with promise for artificial neurons. Both nanowires and Josephson junctions have quantized flux outputs (flux here being defined as the time-integral of the voltage). However, Josephson devices have outputs of only a single flux-quantum, while nanowires have outputs of 50-100 flux quanta (i.e. 70 flux quanta for the device in Figure 2). In pushing the fan-out and fan-in to larger systems, one expects to be eventually limited by parasitic inductances and thermal noises. In such a case, the more substantial signal of the nanowire neuron would permit a larger fan-out and fan-in, leading to a higher degree of parallelism. In addition to the fan-out, the nanowire voltage signals are long

enough and large enough to be digitized directly on an oscilloscope, in contrast to their Josephson junction counterparts, allowing for more direct analysis and readout. Finally, there are a suite of three-terminal, power-control devices made from nanowires [26], which could be fabricated alongside the neurons and used to boost signal and match impedances.

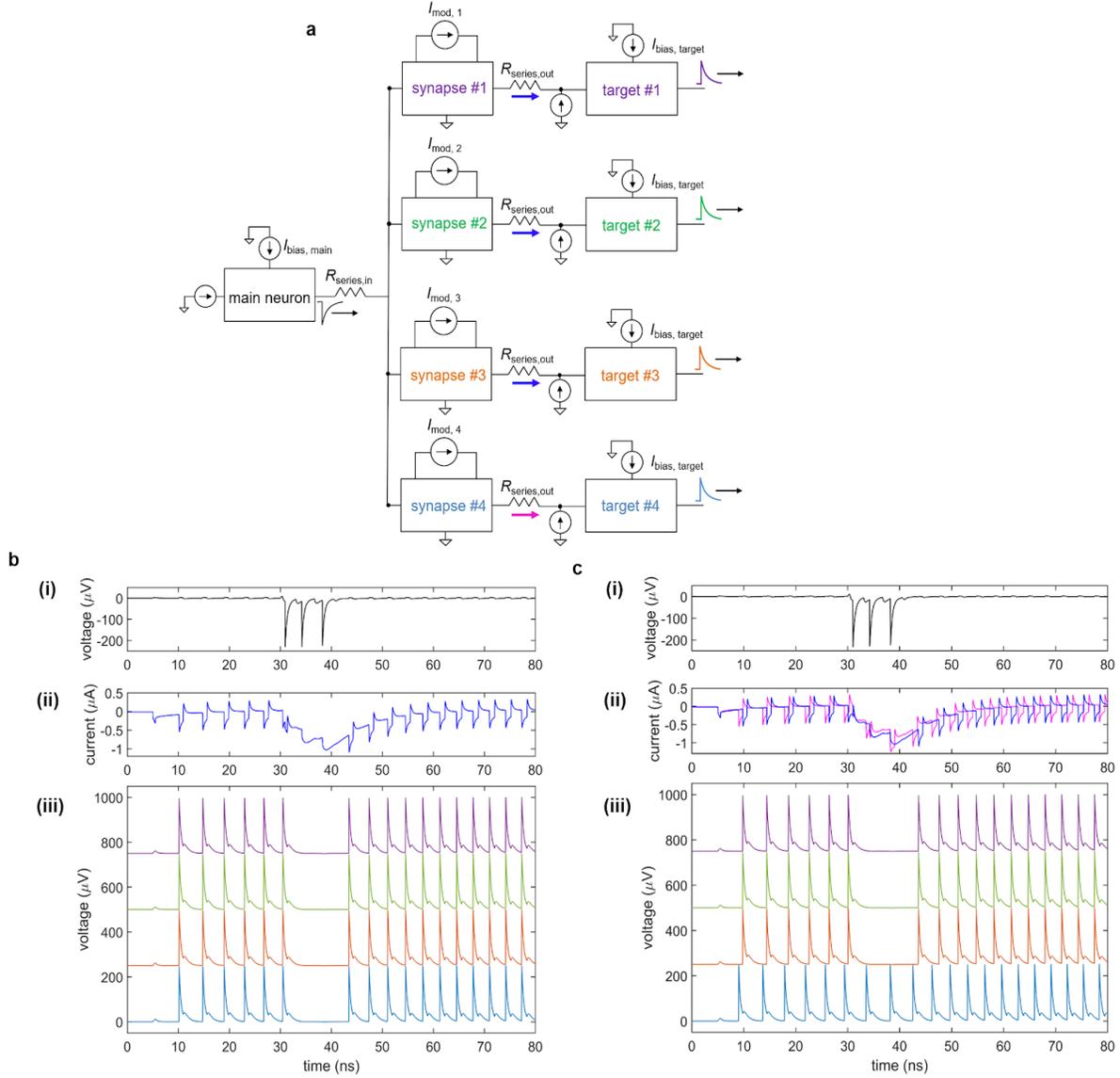

**Figure 12: Fanout of nanowire neuron with a tunable inductive synapse.** (a) Simplified circuit model of the fanout circuit. (b) Simulation when $I_{mod} = 0$ µA for all four target neurons. (i) Output of the main upstream neuron. (ii) Current through each of the four series resistors Rseries out to each of the target neurons. (iii) Output voltage of the four target neurons, shifted on the y-axis for clarity. (c) Simulation when $I_{mod} = 0$ µA for targets 1-3 and $I_{mod} = 8$ µA for target #4. (i) Output of the main upstream neuron. (ii) Current through each of the four series resistors Rseries out to each of the target neurons. The current through $R_{series,out}$ for target #4 is less than that of the other targets, showing that it has been modulated. (iii) Output voltage of the four target neurons, shifted on the y-axis for clarity. The first three targets have been inhibited, while the inhibitory action on target #4 has been turned off. Parameters: $R_{series,in} = 7$ Ω, $R_{syn,1} = R_{syn,2} = 300$ Ω, $L_{syn} = 0.4$ µH, $R_{series,out} = 1$ Ω, $L_{nanowire} = 0.275$ µH, $I_{c,\,nanowire} = 6$ µA, $I_{bias,main} = -59.5$ µA, $I_{bias,target} = 57.65$ µA.

*4.3 Benchmarking Synaptic Energy and Speed*

The energy dissipation of both the neuron and the synapse can be calculated using LTSpice by taking the time integral of the current-voltage product of each circuit element. Performing this analysis, we find that the nanowire neuron has an energy of about 0.05 fJ for each action potential, while the synapse is about an order of magnitude less at around 0.005 fJ. In large systems, it will be the synapses that will dominate; typically if there are $O(N)$ neurons there will be $O(N^2)$ synapses. Hence, even though the neuron dissipates more energy, it will be the synapses that will dominate the power consumption of a large network.

In a spiking neural network, the energy dissipated increases with the speed of the system; spiking twice as often dissipates twice the energy, assuming the energy per spike stays constant. The appropriate figure of merit to compare different technologies is then to take the ratio of speed and power. IBM [27] introduced the figure of merit of synaptic operations per second per watt (SOPS/W). We include a constant factor of about 400 W/W for the nanowire neuron to account for the cryogenic cooling costs. Table 1 compares both the energy per spike and the SOPS/W for the nanowire neuron, the human brain, and two CMOS technologies. We acknowledge that the estimate for the nanowire neuron is a projection from a calculation of a single component, whereas the other entries in the table have actually been measured on large systems. However, given that superconducting platforms have no dissipation in their interconnects, we believe that it is reasonable to project in such a way. In doing so, it is apparent that the nanowire neuron can be a highly competitive technology from a power and speed prospective.

|  | Human brain | NeuroGrid | TrueNorth | Nanowire neuron |
|---|---|---|---|---|
| energy/ switch | 10 fJ | 100 pJ | 25 pJ | ~0.05 fJ |
| SOPS/Watt | 1e14 | 1e10 | 4e10 | 5e14 |

**Table 1: Energy and figure of merit values for different artificial neural networks in comparison to those of the human brain**. Energy values for the human brain, NeuroGrid, and TrueNorth are taken from Ref. [28].

*4.4 Discussion on Possible Synapse Alternatives*

Although the fanout achieved here is far from the level of parallelism in the human brain (where each neuron connects to thousands of neighbors), it suggests that nanowires may serve a unique purpose in the development of future superconducting neural networks, which have thus far struggled to support fanout. Given that nanowires can interface with both CMOS and Josephson junction circuits [29], it may be possible for nanowire neurons to serve as intermediary devices in a network with both platforms. Indeed, a recently proposed neural network with hybrid technologies employed superconducting nanowires as photon detectors, relying instead on optical signals for facilitating high fanout [30] [31]. Although our work uses nanowires solely as electrical components, they can easily be biased to act as photodetectors [32] [33] as well, illustrating that the two different architectures would be compatible for integration. While our work takes more of an electrical approach, the two architectures illustrate the diverse ways in which superconducting nanowires can be used in neural networks, suggesting that a combination of the two technologies may be possible.

Furthermore, while the inductive synapse proposed in this work relies on an external modulating bias current, it may be possible to use a superconducting memory cell to store the value of this modulating current, or replace it directly in the circuit. Memory cells based on superconducting nanowire loops have been shown to reliably store states in the form of a trapped circulating current [34] [35] [36], with recent work demonstrating that one can program and store different amounts of current in the loop in discrete quantities [37]. Incorporating these programmable memory cells into the modulating elements of the inductive synapse could offer multi-level storage of synaptic strengths, allowing for more complex applications. It may also be possible for nanowire neurons to connect through alternative synapse designs, such as one based on inductive coupling. Further analysis is needed to verify the possibility of fanout in such a scheme.

## 5. CONCLUSION

By taking advantage of intrinsic nonlinearities in superconducting nanowires, we have developed a platform for a low-power artificial neuron. In this platform, the coupling of two nanowire-based oscillators acts analogously to the two ion channels in a simplified neuron model, producing an output voltage spike that serves as the information-carrying token in these circuits. Using electrothermal circuit simulations, we have shown that the nanowire neuron is able to reproduce universal biological neuron characteristics, such as a firing threshold, as well as unique characteristics distinct to certain neural classes, such as parabolic bursting. Furthermore, we suggested that a nanowire transmission line with a propagation speed of ~ 2% $c$ may be used as an axon delay line, potentially allowing spatio-temporal information to be accessed. These collective behaviors may enable the nanowire neuron to be used in a rich variety of operations.

In addition to harnessing the nonlinearity of the nanowire's switching dynamics, we relied on the nonlinearity of the nanowire's kinetic inductance in order to develop a variable inductive synapse. We demonstrated that the total parallel inductance of the synapse can be tuned with a modulating bias current, thereby changing the strength of the signal sent to a downstream target neuron. This scheme proved to be capable of fanout, as demonstrated by a single neuron inhibiting the firing of four target neurons. An energy analysis of this circuit in comparison to other spiking networks illustrated that the nanowire neuron has competitive performance in the dynamic firing state and a figure of merit two orders of magnitude better than certain alternative platforms, while the static state benefits from the lack of power dissipation by the superconducting elements. Although experimental realization of these results is a subject of future work, the analysis performed here suggests that the nanowire neuron is a promising candidate for the advancement of low-power artificial neural networks.

Looking forward, networks of superconducting nanowires could be the basis for powerful new computer hardware. Analog blocks of highly connected neurons could be digitally linked to achieve networks with either scale-free or small-worlds connectivity. With superconducting interconnects, entire chips could be wired together with no cost in heat dissipation. The result would be a large-scale neuromorphic processor which could be trained as a spiking neural network to perform tasks like pattern recognition or used to simulate the spiking dynamics of a large, biologically-realistic network. The combination of speed, low power dissipation, and biological realism with only a few components suggests that nanowires could outperform or complement other existing and developing neuromorphic hardware technologies.

**ACKNOWLEDGEMENTS**

The authors thank Murat Onen and Brenden Butters for scientific discussions, Akshay Agarwal and Dr. Donnie Keithley for edits, and the entire Quantum Nanostructures and Nanofabrication group. They would also like to acknowledge the generous support of the Bose Foundation. Emily Toomey was supported by the National Science Foundation Graduate Research Fellowship Program (NSF GRFP) under Grant No. 1122374.

**AUTHOR CONTRIBUTIONS**

E.T. performed the simulations with input from K.S. and K.K.B. All authors contributed to the design of the circuits and the writing of the paper.

**SUPPLEMENTARY MATERIAL**

An example of a capacitive synapse, as well as a demonstration of the tunable inductive synapse in an excitatory case, may be found in the Supplementary Material.

**DATA AVAILABILITY**

The nanowire LTSpice model used for this study can be found on GitHub at https://github.com/karlberggren/snspd-spice. All circuit schematic and data files, including the Matlab code used to generate the figures, are available from the authors upon request.